\documentclass{elsart1p}

\usepackage{graphicx}
\usepackage[bookmarks=true,bookmarksnumbered=true,plainpages=false]{hyperref}

\usepackage{amssymb}

\begin{document}

\begin{frontmatter}



\title{The Fluid Nature of Quark-Gluon Plasma}


\author{W.A. Zajc}
\ead{zajc@nevis.columbia.edu}
%\ead[http://www.nevis.columbia.edu/~zajc/]{home page}

\address{Physics Department\\
Columbia University\\
New York, NY 10027}

\begin{abstract}
Collisions of heavy nuclei at very high energies offer the exciting
possibility of experimentally exploring the phase transformation
from hadronic to partonic degrees of freedom which is predicted to
occur at several times normal nuclear density and/or for
temperatures in excess of $\sim 170$~MeV. Such a state, often
referred to as a quark-gluon plasma, is thought to have been the
dominant form of matter in the universe in the first few
microseconds after the Big Bang. Data from the first five years of
heavy ion collisions of Brookhaven National Laboratory's
Relativistic Heavy Ion Collider (RHIC) clearly demonstrate that
these very high temperatures and densities have been achieved. While
there are  strong suggestions of the role of quark degrees of
freedom in determining the final-state distributions of the produced
matter, there is also compelling evidence that the matter does {\em
not} behave as a quasi-ideal state of free quarks and gluons.
Rather, its behavior is that of a dense fluid with very low
kinematic viscosity exhibiting strong hydrodynamic flow and nearly
complete absorption of high momentum probes. The current status of
the RHIC experimental studies is presented, with a special emphasis
on the fluid properties of the created matter, which may in fact be
the most perfect fluid ever studied in the laboratory.
\end{abstract}

\end{frontmatter}

% main text
\section*{{\bf Preface:}}
\label{Sec:Shoji}
This manuscript is respectfully dedicated to the chair of INPC07, Prof.~Shoji Nagamiya,
for his extraordinary contributions to all of nuclear physics, but especially for his essential
roles at RHIC, on PHENIX, at Columbia and in influencing this author's professional career.
\section{Introduction}
\label{Sec:Intro}
Experiments at Brookhaven National Laboratory's Relativistic Heavy Ion Collider
have achieved their goals of creating and characterizing a new state of matter,
which has come to be known as the strongly-coupled Quark-Gluon Plasma (sQGP).
The striking discoveries and their implications from the initial three years
of RHIC operations were extensively detailed in ``white papers'' from the four
experiments- BRAHMS\cite{Arsene:2004fa}, PHENIX\cite{Adcox:2004mh}, PHOBOS\cite{Back:2004je} and STAR\cite{Adams:2005dq}
which, together with understanding developed in previous theoretical work\cite{Rischke:2005ne},
led to the announcement\cite{PerfectLiquid} of the ``perfect liquid'' behavior of the matter
produced at RHIC.

Since that time, analysis of substantially larger data sets has provided
additional experimental evidence
in support of those statements.
Perhaps more importantly, further discoveries
have produced both new insights and new puzzles which demand more
detailed experimental and theoretical investigation.
Following a brief
review, a sampling of these new experimental results and their implications will be presented.
Due to length restrictions, the topics covered, while adhering closely to those presented
at the conference, will be even more selective
than the author's talk\cite{ZajcTalk}, to which the
reader is referred for supporting material.

\section{The Initial RHIC Discoveries}
\label{Sec:Initial}

During the initial phase of RHIC operations
it was quickly established that the relative abundances and spectra of the particles
produced in Au+Au collisions at RHIC energies (initially $\sqrt{s_{NN}}=130$~GeV
in RHIC Run-1, followed by 200~GeV collisions in RHIC Run-2 in 2001-2) were
consistent with emission from a thermally equilibrated
source\cite{BraunMunzinger:2001ip,Florkowski:2001fp,Cleymans:2004pp,Rafelski:2004dp}
with a chemical freeze-out temperature $T_0 \sim 170$~MeV and a low baryon chemical
potential in a manner consistent with trends seen in lower energy collisions\cite{Andronic:2005yp}.
\begin{figure}
\begin{center}
%\hfill
\includegraphics[height=3.2in]{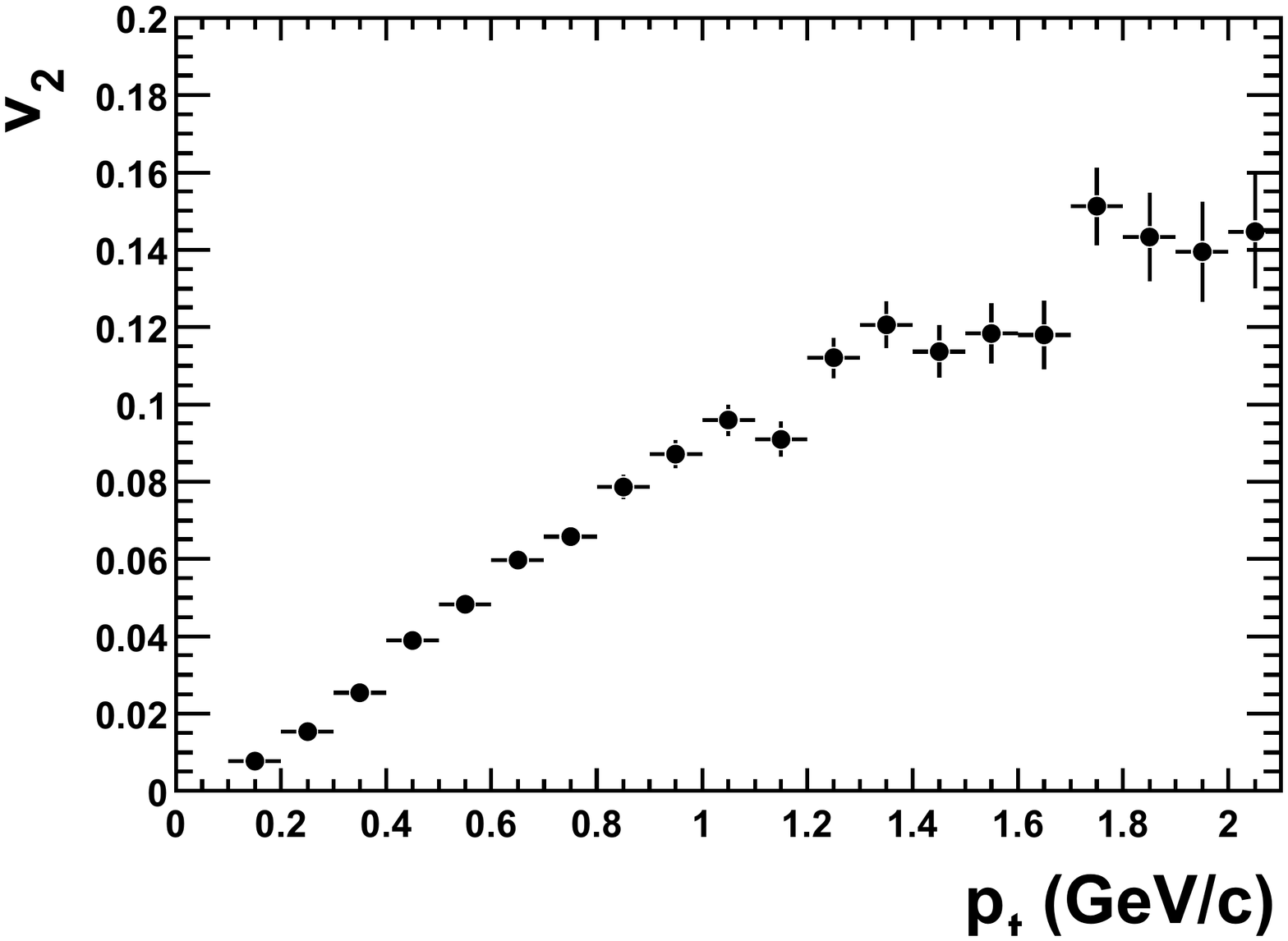}
\includegraphics[height=3.0in]{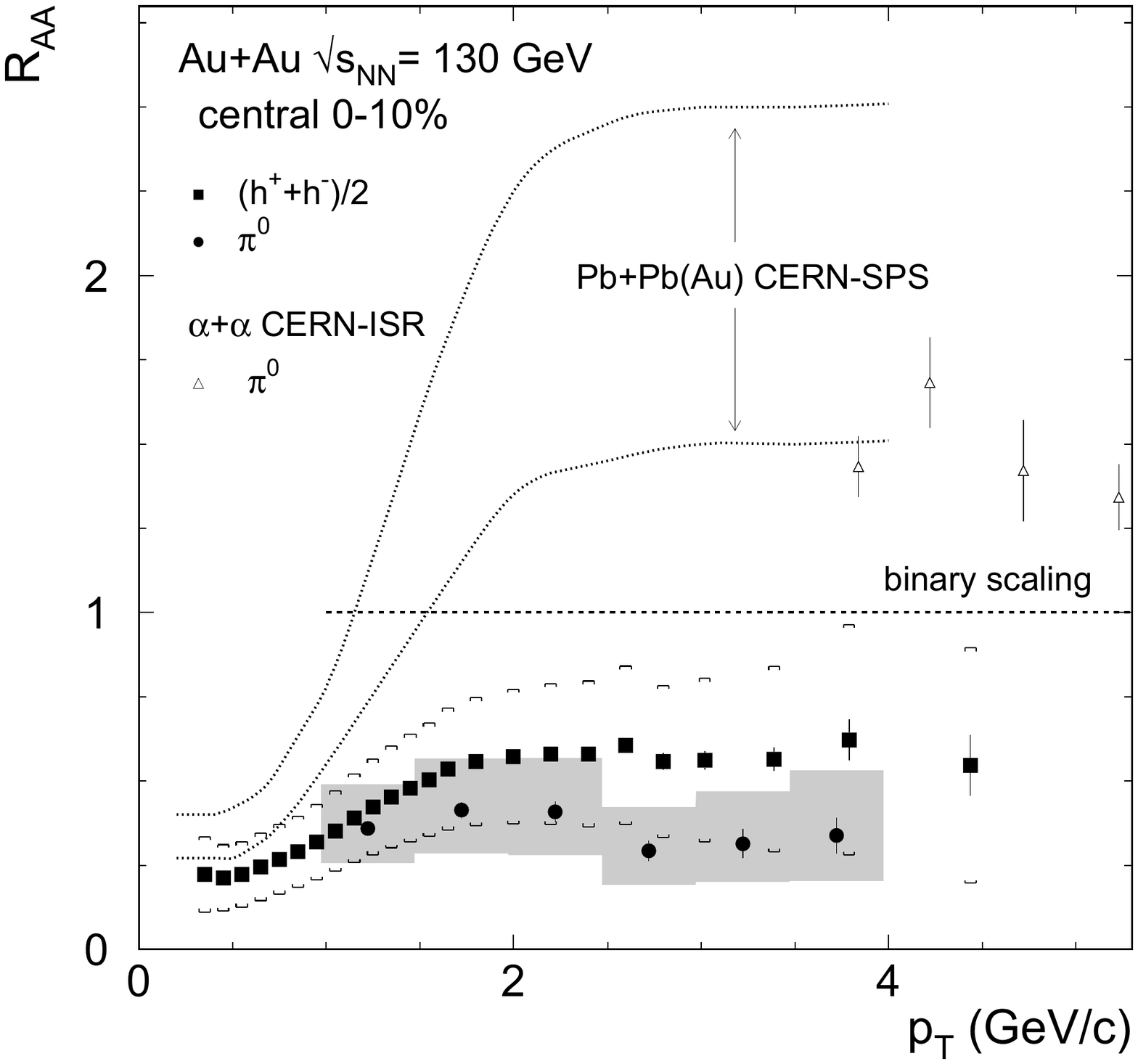}
%\hfill
\end{center}
\caption{The discovery in $\sqrt{s_{NN}}=130~GeV/c$ Au+Au collisions at RHIC of
strong elliptic flow (left, \cite{Ackermann:2000tr}) and of jet quenching (right, \cite{Adcox:2001jp}).
Left: The flow strength parameter $v_2$ versus transverse momentum $p_T$ for
charged particles produced at mid-rapidity in minimum bias collisions.
Right: The suppression factor $R_{AA}$ versus $p_T$ for $\pi^0$'s (circles)
and charged particles (squares) in central collisions, compared to lower
energy results.}
\label{Fig:Discovery}
\end{figure}
However, analysis of the yields as a function of the angle $\phi$
with respect to the reaction plane of the collision revealed the presence of
strong ``elliptic'' flow\cite{Ackermann:2000tr} (Figure~\ref{Fig:Discovery}),
parameterized in terms of the Fourier coefficient $v_2$ in the
expansion
\begin{equation}
{ dn \over { d\phi } } \sim  1 + 2 v_2(p_T) \cos 2\phi + \cdots
\end{equation}
(here $v_2$ is also regarded as a function of the transverse momentum $p_T \equiv |{\vec p}| \sin \theta$
of the emitted particle).
In contrast to behavior at lower energies, the strength of this angular modulation,
and its systematic variation with the
mass of the produced particles, was found for the first time to be consistent with the solutions
of {\em ideal} (non-viscous) hydrodynamics. We will return to the importance of this observation
in Section~\ref{Sec:EtaLimits}.

The discovery of ``jet quenching'' at RHIC\cite{Adcox:2001jp}-
a strong suppression in the production of high $p_T$ particles in nuclear collisions relative
to the expected yield based on  p+p collisions,
also stands in stark contrast to results at lower energy, where particle production at high
transverse momentum is enhanced rather than suppressed in nucleus-nucleus collisions (Figure~\ref{Fig:Discovery}).
Expressed in terms of the ratio $R_{AA}(p_T)$, defined as
\begin{equation}
R_{AA}(p_T) \equiv
{
{\mathrm Yield\ \ in\ \ Au+Au\ \ events}
\over
{\mathrm Scaled\ \ Yield\ \ in\ \  p+p\ \ events}
} \quad ,
\end{equation}
where the denominator consists of the p+p yield scaled, as per {\em
perturbative} QCD (pQCD) by the equivalent parton+parton flux from
a Au+Au collision,
the suppression was found to be as large as a
factor of 5 in the most central events at
$\sqrt{s_{NN}}=200$~GeV\cite{Adler:2003qi,Adams:2003kv}.
In a curious inversion,
the realization\cite{Levai:2001dc} that
detailed information on the opacity and other properties of a dense thermal QCD system
could be obtained using the very deviations from pQCD expectations {\em
absent interactions in a produced medium}
spurred development and application
of a sophisticated
technology\cite{Baier:1996kr,Zakharov:1997uu,Wiedemann:2000za,Baier:2001yt,Gyulassy:2000er,Gyulassy:2002yv}
making possible ``tomographic'' studies of the produced matter.
The observed quenching was consistent with parton energy loss rates
$\sim15$ times higher than in cold nuclear matter\cite{Wang:2002ri},
and demanded an initial matter density of order 100 times that of
normal nuclear matter\cite{Gyulassy:2001kr,Gyulassy:2001nm,Salgado:2002cd}.
\begin{figure}
\begin{center}
%\hfill
\includegraphics[height=3.5in]{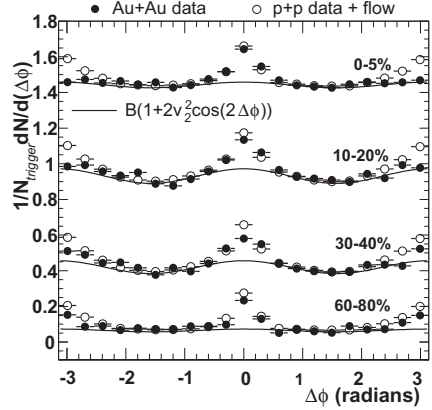}
%\hfill
\end{center}
\caption{Away-side jet disappearance: the angular distribution of high $p_T$ ($> 2$~GeV/c) particles
associated with a trigger particle $p_T^{trig} \in [4,6]$~GeV/c measured
in p+p (open circles) and Au+Au (filled circles) collisions as a function
of centrality (expressed as the percentage of the total cross section).
Taken from Ref.~\cite{Adler:2002tq}.}
\label{Fig:STARdisappear}
\end{figure}
A striking observation in support of these estimates was the
disappearance of the ``away-side'' jet partner in Au+Au collisions\cite{Adler:2002tq}~(Figure~\ref{Fig:STARdisappear}),
indicating that the matter density was essentially opaque to high-$p_T$ partons
and that the observed high transverse momentum ``trigger'' particles were
dominated by surface emission.

Three other early key developments can only be briefly mentioned here:
\begin{itemize}

\item The interpretation of the jet-quenching results was bolstered
by reliance on {\em in situ} measurement of baseline (p+p) and control (d+Au)
data. Comparison of the p+p data to theoretical calculations established the
quantitative reliability of pQCD calculations at RHIC energies\cite{Adler:2003pb}.
The demonstration that suppression effects were
absent in d+Au collisions\cite{Back:2003ns,Adler:2003ii,Adams:2003im,Arsene:2003yk}
provided crucial evidence that the quenching observed in Au+Au collisions
was due to parton propagation in a dense thermal environment, rather than
to modifications of the nuclear wave function.

\item While both thermal and hydrodynamic models worked with unprecedented
accuracy to describe the final state distribution of particles at RHIC,
the actual abundances were low compared to pre-RHIC expectations\cite{Bass:1999zq}.
Both the low value of the charged-particle density\cite{Back:2000gw} and its systematic variation
with centrality and energy\cite{Adcox:2000sp,Back:2001ae,Adler:2004zn}
were described quantitatively in a model incorporating gluon saturation
in the initial nuclear wave-function\cite{Kharzeev:2001yq}.
Subsequent developments have shown the importance of this observation
for establishing reliable initial conditions for hydrodynamic
calculations at RHIC energies\cite{Hirano:2004en}.

\item The surprising observation that baryon and anti-baryon yields
were comparable to those of mesons at intermediate (4-6~GeV/c) transverse
momenta\cite{Adcox:2001mf} was not understandable in terms of
standard thermal production models with a chemical freeze-out condition.
However, this so-called ``baryon anomaly''  found a natural explanation
in recombination models\cite{Hwa:2002tu,Fries:2003vb,Greco:2003xt,Greco:2003mm,Fries:2003kq},
which in a sufficiently dense partonic medium will favor the creation of final-state
hadrons via {\em coalescence} of lower-momentum quarks drawn from a thermal spectrum over the
{\em fragmentation} of a higher-momentum parton to a hadron with lower momentum.
The successes of this mechanism, which clearly requires partonic degrees of freedom
in a dense medium, led the authors of Ref.~\cite{Fries:2003vb} to note
`` \dots our scenario requires the assumption of a thermalized partonic phase characterized
by an exponential momentum spectrum. Such a phase may be appropriately called a quark-gluon plasma.''
\end{itemize}

\section{More Recent Developments}
The steady improvement in RHIC
luminosity\footnote{RHIC now routinely operates at more
than 4 times its design luminosity of $2\times10^{26} \mathrm {cm^{-2}s^{-1}}$
for Au+Au collisions.}
and in experimental data-taking
capabilities has led to an increase approaching
three orders of magnitude in integrated luminosity over
that of the initial (Run-1) discovery period. It is not possible
to enumerate, much less fully explore, all of the results
obtained from this cornucopia of data.
(See Ref.~\cite{Muller:2006ee} for a recent review.)
Instead, focus will be applied to a very limited subset
of topics most relevant to future investigations that seek
to characterize the properties of the medium.

\subsection{Direct Photon Measurements}
While the absence of suppression effects observed in d+Au
collisions also suggested that the pQCD scaling methodology
was well-controlled, it obviously did not directly demonstrate that
this was also the case for Au+Au collisions. High energy
($p_T > \sim 3$~GeV/c) direct photons, while experimentally
very difficult to separate from the copious background from
$\pi^0$ and $\eta$ decays, do provide the desired calibration,
since their production rate is proportional to the initial
parton+parton flux and they interact only very weakly with
medium. The measurement of $R_{AA}$ for direct photons
in Au+Au collisions\cite{Adler:2005ig}, which requires tight control of experimental
systematics over several orders of magnitude, clearly
establishes the  validity of the assumed scaling techniques.

The effort now turns to ever-increasing precision in
measuring and normalizing the observed photon yield,
in search of predicted $\sim20$\% effects due to the interplay
of isospin, fragmentation, shadowing and energy loss\cite{Turbide:2005fk,Arleo:2006xb}.
An important experimental development is the first
proof-of-principle results on gamma-hadron correlations\cite{Jin:2007by,Chattopadhyay:2007in},
which are a necessary precursor to the long-desired goal of
using direct photons as a calibrated tag to measure precisely
jet energy loss in nuclear collisions\cite{Wang:1996yh}.

\subsection{Detailed Investigations of Hydrodynamic Behavior}
The consistency of the RHIC experimental data on elliptic flow with
calculations based on ideal hydrodynamics, together with the jet
quenching results demonstrating the extraordinary density of the
matter, resulted in the descriptor ``perfect liquid'', in analogy
with usage of  ``perfect fluid'' in general relativity to denote a
fluid that is completely isotropic to co-moving
observers\cite{Weinberg} (thereby implying zero viscosity and
perfect heat conductivity). A great deal of experimental and
theoretical work is underway to determine the kinematic regime in
which this description (approximately) applies, and to quantify the
transport properties of the near-perfect medium. A highly restricted
sample of these efforts is presented in this section.
\subsubsection{Scaling Behavior of Elliptic Flow}
As noted in Section~\ref{Sec:Initial}, data from RHIC show that the detailed dependence of
the elliptic flow parameter $v_2(p_T)$ on particle mass
is consistent with calculations based on ideal hydrodynamics. This observation
has been considerably sharpened by the discovery of a family of
exact scaling solutions to ideal hydrodynamics\cite{Csanad:2006sp} which
show a universal scaling behavior of the elliptic flow parameter $v_2$
upon a reduced kinematic variable. Near mid-rapidity, this variable
reduces (approximately) to ``transverse kinematic energy''
$ {\mathrm KE_T} \equiv \sqrt{m^2+p_T^2}-m$.
\begin{figure}
\begin{center}
%\hfill
\includegraphics[height=3.0in]{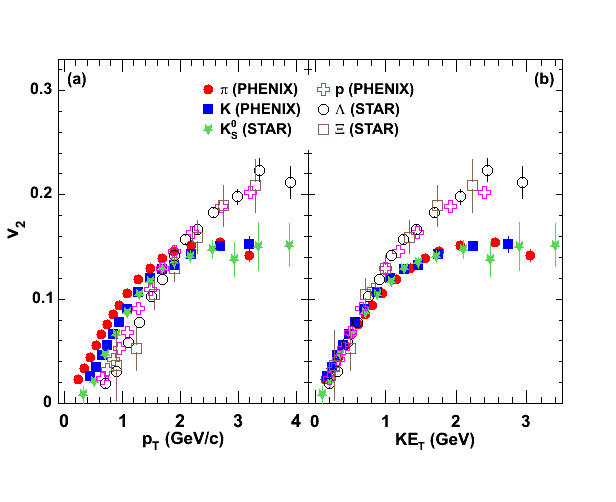}
%\hfill
\end{center}
\caption{The scaling\cite{Adare:2006ti} with
transverse kinetic energy $\mathrm KE_T$ of PHENIX and STAR\cite{Adams:2003am,Adams:2005zg}
data for the elliptic flow parameter $v_2(p_T)$ for various particles.}
\label{Fig:KETScaling}
\end{figure}
Indeed, the mass-dependent structure in $v_2(p_T)$ does fall on such a
scaling curve\cite{Adare:2006ti}, as shown in
Figure~\ref{Fig:KETScaling}. While the scaling behavior appears limited to
$\mathrm KE_T < ~1.0$~GeV, note that a)~more than 98\% of the produced particles are
in this domain and b)~the larger values of $\mathrm KE_T$ correspond to a regime
where hard scattering may already play a role,
potentially explaining the observed deviations from hydrodynamic behavior (however, on this point
see further discussion in Sections \ref{Sec:EtaLimits} and \ref{Sec:Mach}).
With these mild caveats, the scaling behavior shown in Figure~\ref{Fig:KETScaling}
is consistent with a key prediction of ideal (inviscid) hydrodynamics for
bulk particle production at RHIC.

\subsection{Bounding Perfection}
\label{Sec:AdS}
Taken at face value, the good agreement between the experimental data
and inviscid hydrodynamics supports Landau's observation\cite{Belenkij:1956cd} that
the very conditions for the applicability of hydrodynamics to nuclear systems
(a mean free path much smaller than the system size) necessarily lead to a
negligible viscosity. One of the most fascinating developments in recent years
has been the conjecture\cite{Kovtun:2004de} that there may be a fundamental bound from below on
the value of viscosity, that is, for any thermal fluid, the uncertainty principle
requires a non-zero viscosity. More precisely, the conjecture states that the
ratio of viscosity $\eta$ to entropy density $s$ must satisfy
\begin{equation}
{\eta \over s} \geq {\hbar \over {4\pi}} \quad .
\label{Eq:Bound}
\end{equation}

Particularly intriguing is the origin of the bound. While the simple dependence
on $\hbar$ implies it is a strictly quantum mechanical result, and while
estimates based on uncertainty relations have been made\cite{Danielewicz:1984ww,Hirano:2005wx}, the first explicit
derivation of a numerical value for the bound was obtained for a
maximally supersymmetric Yang-Mills theory via the AdS/CFT correspondence\cite{Maldacena:1997re}.
By exploiting a duality between (string) quantum gravity in a higher-dimensional Anti de Sitter(AdS) spacetime
and conformal field theory (CFT) on the boundary of AdS, strongly coupled problems in the field theory
are mapped onto weakly coupled, and thus semi-classical, gravity calculations in the bulk.
As a result, the entropy density in the field theory is dual to the entropy of extended black branes,
and the viscosity is dual to graviton absorption by the brane. While the supersymmetric conformal field theory
would appear to be far removed from non-conformal non-supersymmetric ordinary QCD, arguments have been made
that for thermal QCD systems somewhat above the critical temperature may at least qualitatively
be regarded as a conformal theory in which the only dimensionful quantity is the temperature\cite{Gubser:2006bz,Liu:2006he}.

\subsection{Approach to Perfection}
\label{Sec:EtaLimits}
Given the (conjectured) existence of a bound on $\eta/s$, it is only natural to
ask how closely does the QGP fluid produced at RHIC approach the bound.
The most natural approach is not to separately extract $\eta$ and $s$
from the data, but instead to note that damping (of sound, flow, etc.),
at temperature $T$ and zero chemical potential is proportional to $\eta \over {s T}$.
This was the basis for a first schematic calculation\cite{Teaney:2003kp} that was instrumental
in developing the case for ideal fluid behavior at RHIC\cite{Shuryak:2003xe}.
Since then, somewhat more sophisticated analyses coupled with greatly improved data
sets have led to first attempts to determine the range of allowed values for
$\eta/s$ in $\sqrt{s_{NN}}=200$~GeV Au+Au collisions at RHIC. Estimates based
on the magnitude of elliptic flow indicate
$\eta/s = (1.1 \pm 0.2 \pm 1.2){1\over{4\pi}}$\cite{Lacey:2006bc}
and
$\eta/s = (1.9      -      2.5){1\over{4\pi}}$\cite{Drescher:2007cd}
,
an analysis of the damping of $p_T$ fluctuations gives
$\eta/s = (1.0      -      3.8){1\over{4\pi}}$\cite{Gavin:2006xd},
while a simultaneous analysis of the observed energy loss and flow of
heavy quarks produces
$\eta/s = (1.3      -      2.0){1\over{4\pi}}$\cite{Adare:2006nq}.
These different methods provide consistent support for the conclusion
that the produced matter is within a factor of 2-3 of the conjectured
bound, that is, has a viscosity to entropy density ratio lower than
any other known fluid. The above values are comparable to, although
generally smaller than, the value $\eta/s \sim 0.5 \sim 6 \ ( {1\over{4\pi}} ) $
obtained from a study of the breathing modes of a gas of cold trapped ${}^6\rm Li$
atoms at the unitary limit\cite{Schafer:2007waz}, which exhibits a viscosity
to entropy density ratio even lower than liquid helium near the lambda point\cite{Kovtun:2004de}.

\subsection{Reaction of the Medium}
\label{Sec:Mach}

\begin{figure}
\begin{center}
%\hfill
%\includegraphics[height=2.5in]{PPG034Polar.png}
\includegraphics[height=2.5in]{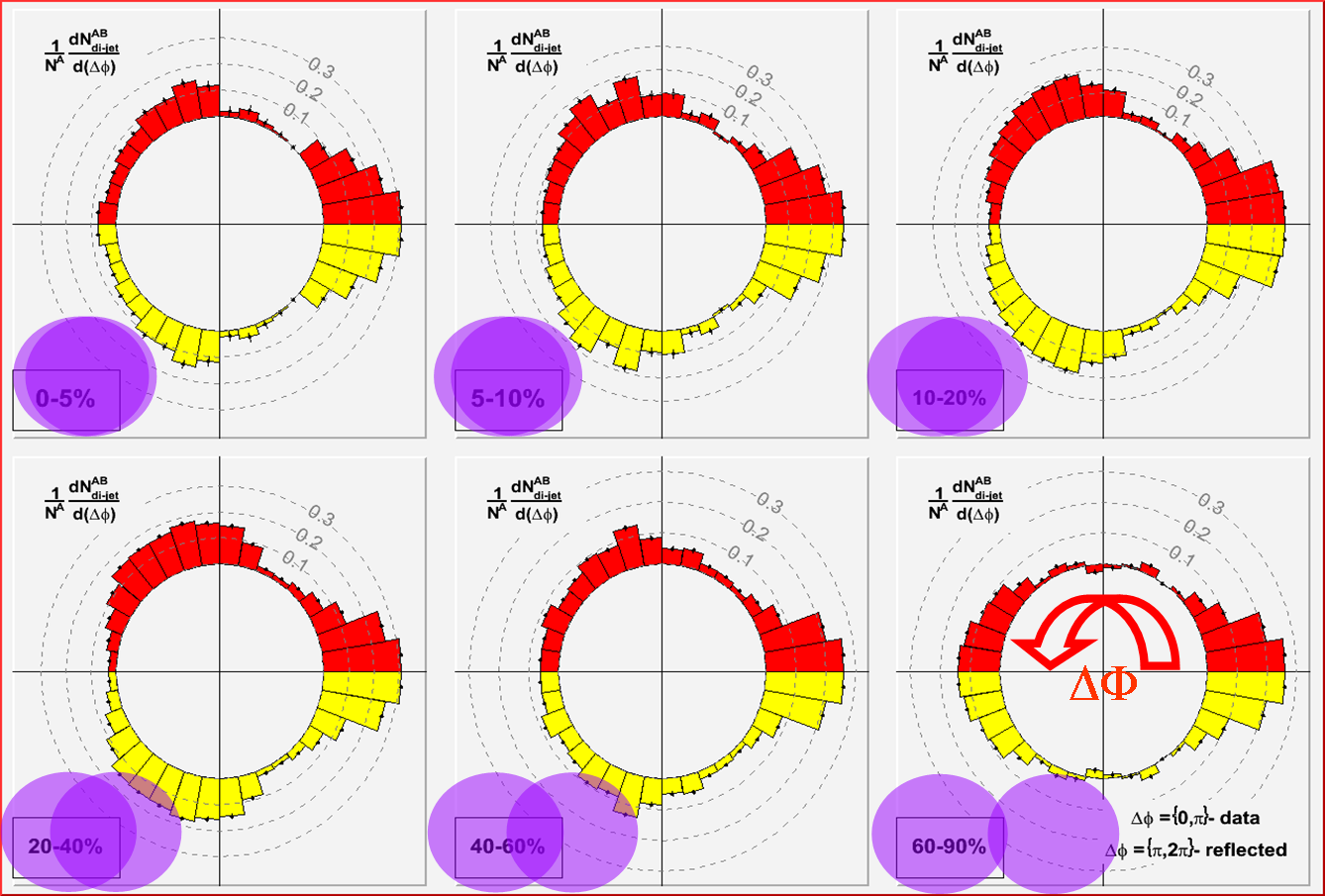}
%\hfill
\end{center}
\caption{A polar representation of the data from Ref.~\cite{Adler:2005ee},
showing the development of strongly modified away-side jet response as
a function of centrality.}
\label{Fig:PPG034Polar}
\end{figure}
While the jet energy loss phenomena would appear to be strictly in the domain
of perturbative QCD, and therefore seemingly divorced from the bulk flow
behavior at low transverse momenta, a fascinating connection between these
effects has emerged in recent studies\cite{Shuryak:2006ii}.
The disappearance of away-side jets shown in Figure~\ref{Fig:STARdisappear}
strongly suggests that the {\em directed} energy in a high momentum
transfer parton scatter is absorbed by the {\em medium}. A sufficiently
dense and strongly coupled medium, while thermalizing that energy, must nonetheless
conserve momentum. One method for doing so is via the development of a
shock or ``Mach'' cone\cite{Stoecker:2004qu,CasalderreySolana:2004qm}, i.e.,
the transfer of the directed energy of an away-side jet into a collective
excitation of lower momentum particles in the medium. Indeed, after
carefully taking into account the modulation of the background due to
elliptic flow, both published data\cite{Adler:2005ee,Adare:2006nr}
and more recent preliminary results\cite{Horner:2007gt,Ulery:2007zb,Jia:2007sf}
provide support for this hypothesis from an examination of the angular correlations
of away-side particles in the range $ \sim 1 \ \rm GeV/c < p_T < \sim 2 \ \rm GeV/c$.

Further work, both experimental and theoretical, is required to determine if the
clearly observed away-side jet distortions at low $p_T$ are in fact Mach cones.
Experimentally, more detailed analysis of both 2 and 3-particle correlations in more
extensive data sets is underway. Theoretically, moving from a bulk description of
hydrodynamic shock fronts to a more microscopic view of the energy-momentum
transport is very challenging. Here too AdS/CFT methods have played a role,
as the duality permits calculations at all length scales.
While this approach suffers not only from the standard concern that the gauge theory
studied in not QCD but also from the restriction to calculating wakes from infinitely massive quarks,
it has nonetheless provided substantial insight into the energy flow and medium response\cite{Gubser:2007ni}.

\subsection{Puzzles from Quark Number Scaling}

\begin{figure}
\begin{center}
%\hfill
\includegraphics[height=2.5in]{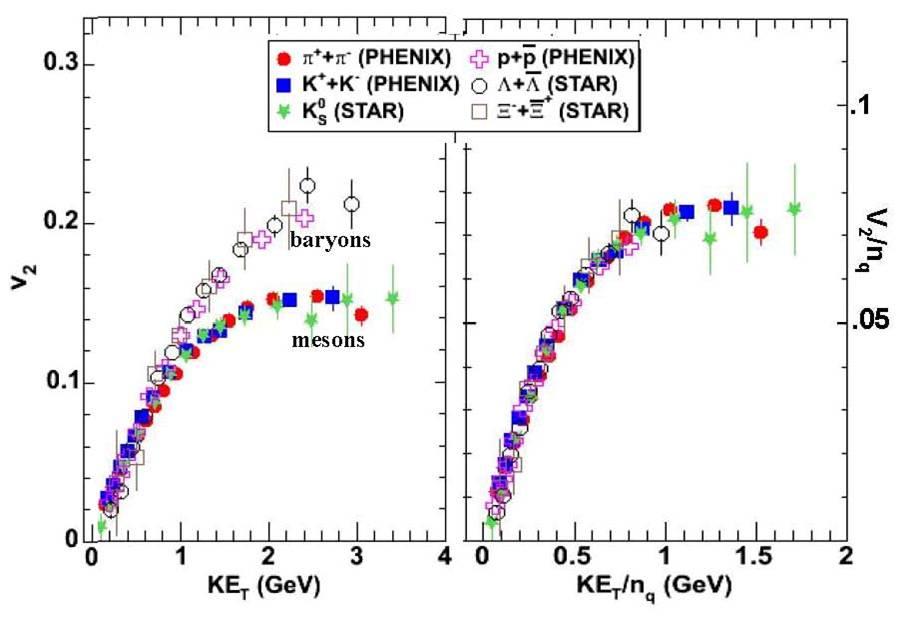}
%\hfill
\end{center}
\caption{The additional scaling of $v_2(\mathrm KE_T)$ with the
number of constituent quarks $n_q$\cite{Adare:2006ti}.}
\label{Fig:nqScaling}
\end{figure}
The recent spectacular advances in our understanding of hot QCD have been driven by experimental discoveries,
together with the subsequent development of detailed theoretical understanding. It is therefore
appropriate to discuss briefly at least one area where such understanding remains elusive.
A case in point is provided by the observed quark number scaling of elliptic flow\cite{Adare:2006ti,Adams:2004bi}.
Examination of the two branches of $v_2(\rm KE_T)$ displayed in Figure~\ref{Fig:KETScaling} shows that
the upper branch consists of baryons; the lower branch mesons. Upon further scaling of both
$\rm KE_T$ and $v_2(\rm KE_T)$ by $n_q$, the number of constituent quarks, the two branches merge
into a universal scaling curve for elliptic flow (Figure~\ref{Fig:nqScaling}).
A critical test of this observation is the $\phi$ meson- although more massive than
a nucleon, it follows the scaling curve for mesons, not baryons\cite{Afanasiev:2007tv,Abelev:2007rw}.
While it is tempting to conclude that the strict scaling according to constituent quark content
provides incontrovertible evidence
for the underlying role of quark degrees of freedom in establishing the elliptic flow,
such a conclusion appears to be at odds with the observation of perfect fluidity.
Quasi-particles with lifetimes comparable to the relatively long time over which hydrodynamic flow persists
are incompatible with the very short mean free paths (and hence large widths and short lifetimes) implied by low viscosity\cite{QP}.
This fact, already noted in the earliest calculation of viscosity via the AdS/CFT correspondence\cite{Policastro:2001yc},
has yet to be reconciled with the strong empirical evidence for quark number scaling of flow phenomena.
Recent calculations\cite{Song:2007fn} which indicate that the viscosity at RHIC may be even lower than
the conjectured bound only sharpen the dichotomy.

\section{Outlook}
The RHIC experiments have conclusively demonstrated the feasibility,
and indeed the desirability, of performing sensitive measurements of
heavy ion phenomena in a collider environment. The discoveries
made in the initial years of RHIC operation have initiated a paradigm
shift in our understanding of deconfined QCD matter in the region of
the transition temperature. In doing so they have also set the stage for
the heavy ion program at the LHC, where Pb+Pb collisions at
$\sqrt{s_{NN}}=5.5$~TeV are planned following heavy ion commissioning in 2009.
This unprecedented increase in center-of-mass energy will undoubtedly result
in a new round of discoveries, to be made by the ALICE experiment\cite{Antinori:2007dna},
which is dedicated to the heavy ion program, together with the two large
p+p collider detectors ATLAS\cite{Steinberg:2007nm} and CMS\cite{Betts:2007zz}.
Questions of immediate interest are the fundamental properties of the QGP at the
expected higher initial temperatures at the LHC- Will it remain strongly coupled?
Will it exhibit $\sim$perfect flow? However, the RHIC experience has taught
us that the very nature of the important questions changes in response to
experimental data, and it would be wise to anticipate that the same will
be true for heavy ion physics at LHC energies.

A compelling program of more detailed investigations at RHIC is already
underway, based on a series of detector upgrades to PHENIX and STAR, together
with ongoing and future substantial increases in RHIC luminosity.
The unique properties of bulk, thermalized QCD matter at RHIC energies
will be better quantified via increasingly differential measurements,
such as associated particle production in direct photon events,
jet tomography with respect to the reaction plane, and
the suppression and flow patterns of both charmonium and open heavy flavor.

A particular strength of the RHIC facility is the wide range of collision energies,
from the top energy of $\sqrt{s_{NN}}=200$~GeV to values as low as a few GeV.
This, together with the collider geometry which provides excellent control
of acceptance systematics as the center-of-mass energy is varied, have stimulated
interest in the search for the QCD critical point\cite{Stephanov:1998dy} at RHIC\cite{Stephans:2006tg}.
Especially intriguing in this regard is the possible connection\cite{Csernai:2006zz} between physics in the region
of the critical point and the minimal value of the viscosity to entropy density ratio.
The simultaneous progress on multiple energy frontiers at RHIC, at the LHC
and at GSI/FAIR\cite{Henning:2004vk}
are certain to lead to new insights and a deeper understanding
of the fluid nature of quark-gluon plasma.

% \appendix

% \section{}
% \label{}


\begin{thebibliography}{00}


%\cite{Arsene:2004fa}
\bibitem{Arsene:2004fa}
  I.~Arsene {\it et al.}  [BRAHMS Collaboration],
  %``Quark gluon plasma and color glass condensate at RHIC? The perspective
  %from the BRAHMS experiment,''
  Nucl.\ Phys.\  A {\bf 757}, 1 (2005)
  [arXiv:nucl-ex/0410020].
  %%CITATION = NUPHA,A757,1;%%

%\cite{Adcox:2004mh}
\bibitem{Adcox:2004mh}
  K.~Adcox {\it et al.}  [PHENIX Collaboration],
  %``Formation of dense partonic matter in relativistic nucleus nucleus
  %collisions at RHIC: Experimental evaluation by the PHENIX  collaboration,''
  Nucl.\ Phys.\  A {\bf 757}, 184 (2005)
  [arXiv:nucl-ex/0410003].
  %%CITATION = NUPHA,A757,184;%%

%\cite{Back:2004je}
\bibitem{Back:2004je}
  B.~B.~Back {\it et al.} [PHOBOS Collaboration],
  %``The PHOBOS perspective on discoveries at RHIC,''
  Nucl.\ Phys.\  A {\bf 757}, 28 (2005)
  [arXiv:nucl-ex/0410022].
  %%CITATION = NUPHA,A757,28;%%


%\cite{Adams:2005dq}
\bibitem{Adams:2005dq}
  J.~Adams {\it et al.}  [STAR Collaboration],
  %``Experimental and theoretical challenges in the search for the quark  gluon
  %plasma: The STAR collaboration's critical assessment of the  evidence from
  %RHIC collisions,''
  Nucl.\ Phys.\  A {\bf 757}, 102 (2005)
  [arXiv:nucl-ex/0501009].
  %%CITATION = NUPHA,A757,102;%%

%\cite{Rischke:2005ne}
\bibitem{Rischke:2005ne}
  D.~Rischke and G.~Levin,
  %``Quark gluon plasma. New discoveries at RHIC: A case of strongly interacting
  %quark gluon plasma. Proceedings, RBRC Workshop, Brookhaven, Upton, USA, May
  %14-15, 2004,''
%\href{http://www.slac.stanford.edu/spires/find/hep/www?irn=6315895}{SPIRES entry}
{\it Prepared for Workshop on New Discoveries at RHIC: The Current
Case for the Strongly Interactive QGP, Brookhaven, Upton, New York,
14-15 May 2004}


%\cite{Lee:2005gw}
%``The strongly interacting quark-gluon plasma and future physics,''
%  Nucl.\ Phys.\  A {\bf 750}, 1 (2005).
%  %%CITATION = NUPHA,A750,1;%%

%\cite{Gyulassy:2004zy}
%\bibitem{Gyulassy:2004zy}
%  M.~Gyulassy and L.~McLerran,
%  %``New forms of QCD matter discovered at RHIC,''
%  Nucl.\ Phys.\  A {\bf 750}, 30 (2005)
%  [arXiv:nucl-th/0405013].
%  %%CITATION = NUPHA,A750,30;%%


\bibitem{PerfectLiquid}
{\em ``RHIC Scientists Serve Up 'Perfect' Liquid''},
\url{http://www.bnl.gov/bnlweb/pubaf/pr/PR\_display.asp?prID=05-38}
.

\bibitem{ZajcTalk}
{\em ``Quark-Gluon Plasma: Experimental Overview''}, W.A.~Zajc,
presented at the International Nuclear Physics Conference; June 3-8,
2007; Tokyo, Japan; available as
\url{http://inpc2007.riken.jp/P/P7-zajc.ppt} .


%\cite{BraunMunzinger:2001ip}
\bibitem{BraunMunzinger:2001ip}
  P.~Braun-Munzinger, D.~Magestro, K.~Redlich and J.~Stachel,
  %``Hadron production in Au Au collisions at RHIC,''
  Phys.\ Lett.\  B {\bf 518}, 41 (2001)
  [arXiv:hep-ph/0105229].
  %%CITATION = PHLTA,B518,41;%%


%\cite{Florkowski:2001fp}
\bibitem{Florkowski:2001fp}
  W.~Florkowski, W.~Broniowski and M.~Michalec,
  %``Thermal analysis of particle ratios and p(T) spectra at RHIC,''
  Acta Phys.\ Polon.\  B {\bf 33}, 761 (2002)
  [arXiv:nucl-th/0106009].
  %%CITATION = APPOA,B33,761;%%

%\cite{Cleymans:2004pp}
\bibitem{Cleymans:2004pp}
  J.~Cleymans, B.~Kampfer, M.~Kaneta, S.~Wheaton and N.~Xu,
  %``Centrality dependence of thermal parameters deduced from hadron
  %multiplicities in Au + Au collisions at s(NN)**(1/2) = 130-GeV,''
  Phys.\ Rev.\  C {\bf 71}, 054901 (2005)
  [arXiv:hep-ph/0409071].
  %%CITATION = PHRVA,C71,054901;%%

%\cite{Rafelski:2004dp}
\bibitem{Rafelski:2004dp}
  J.~Rafelski, J.~Letessier and G.~Torrieri,
  %``Centrality dependence of bulk fireball properties at RHIC,''
  Phys.\ Rev.\  C {\bf 72}, 024905 (2005)
  [arXiv:nucl-th/0412072].
  %%CITATION = PHRVA,C72,024905;%%


%\cite{Andronic:2005yp}
\bibitem{Andronic:2005yp}
  A.~Andronic, P.~Braun-Munzinger and J.~Stachel,
  %``Hadron production in central nucleus nucleus collisions at chemical
  %freeze-out,''
  Nucl.\ Phys.\  A {\bf 772}, 167 (2006)
  [arXiv:nucl-th/0511071].
  %%CITATION = NUPHA,A772,167;%%


%\cite{Ackermann:2000tr}
\bibitem{Ackermann:2000tr}
  K.~H.~Ackermann {\it et al.}  [STAR Collaboration],
  %``Elliptic flow in Au + Au collisions at s(N N)**(1/2) = 130-GeV,''
  Phys.\ Rev.\ Lett.\  {\bf 86}, 402 (2001)
  [arXiv:nucl-ex/0009011].
  %%CITATION = PRLTA,86,402;%%

%\cite{Adcox:2001jp}
\bibitem{Adcox:2001jp}
  K.~Adcox {\it et al.}  [PHENIX Collaboration],
  %``Suppression of hadrons with large transverse momentum in central  Au + Au
  %collisions at s**(1/2)(N N) = 130-GeV,''
  Phys.\ Rev.\ Lett.\  {\bf 88}, 022301 (2002)
  [arXiv:nucl-ex/0109003].
  %%CITATION = PRLTA,88,022301;%%

%\cite{Adler:2003qi}
\bibitem{Adler:2003qi}
  S.~S.~Adler {\it et al.}  [PHENIX Collaboration],
  %``Suppressed pi0 production at large transverse momentum in central Au +  Au
  %collisions at s(NN)**(1/2) = 200-GeV,''
  Phys.\ Rev.\ Lett.\  {\bf 91}, 072301 (2003)
  [arXiv:nucl-ex/0304022].
  %%CITATION = PRLTA,91,072301;%%

%\cite{Adams:2003kv}
\bibitem{Adams:2003kv}
  J.~Adams {\it et al.}  [STAR Collaboration],
  %``Transverse momentum and collision energy dependence of high p(T) hadron
  %suppression in Au + Au collisions at ultrarelativistic energies,''
  Phys.\ Rev.\ Lett.\  {\bf 91}, 172302 (2003)
  [arXiv:nucl-ex/0305015].
  %%CITATION = PRLTA,91,172302;%%




%\cite{Levai:2001dc}
\bibitem{Levai:2001dc}
  P.~Levai, G.~Papp, G.~I.~Fai, M.~Gyulassy, G.~G.~Barnafoldi, I.~Vitev and Y.~Zhang,
  %``Discovery of jet quenching at RHIC and the opacity of the produced  gluon
  %plasma,''
  Nucl.\ Phys.\  A {\bf 698}, 631 (2002)
  [arXiv:nucl-th/0104035].
  %%CITATION = NUPHA,A698,631;%%

%\cite{Baier:1996kr}
\bibitem{Baier:1996kr}
  R.~Baier, Y.~L.~Dokshitzer, A.~H.~Mueller, S.~Peigne and D.~Schiff,
  %``Radiative energy loss of high energy quarks and gluons in a  finite-volume
  %quark-gluon plasma,''
  Nucl.\ Phys.\  B {\bf 483}, 291 (1997)
  [arXiv:hep-ph/9607355].
  %%CITATION = NUPHA,B483,291;%%

%\cite{Zakharov:1997uu}
\bibitem{Zakharov:1997uu}
  B.~G.~Zakharov,
  %``Radiative energy loss of high energy quarks in finite-size nuclear  matter
  %and quark-gluon plasma,''
  JETP Lett.\  {\bf 65}, 615 (1997)
  [arXiv:hep-ph/9704255].
  %%CITATION = JTPLA,65,615;%%


%\cite{Wiedemann:2000za}
\bibitem{Wiedemann:2000za}
  U.~A.~Wiedemann,
  %``Gluon radiation off hard quarks in a nuclear environment: Opacity
  %expansion,''
  Nucl.\ Phys.\  B {\bf 588}, 303 (2000)
  [arXiv:hep-ph/0005129].
  %%CITATION = NUPHA,B588,303;%%


%\cite{Baier:2001yt}
\bibitem{Baier:2001yt}
  R.~Baier, Y.~L.~Dokshitzer, A.~H.~Mueller and D.~Schiff,
  %``Quenching of hadron spectra in media,''
  JHEP {\bf 0109}, 033 (2001)
  [arXiv:hep-ph/0106347].
  %%CITATION = JHEPA,0109,033;%%

%\cite{Gyulassy:2000er}
\bibitem{Gyulassy:2000er}
  M.~Gyulassy, P.~Levai and I.~Vitev,
  %``Reaction operator approach to non-Abelian energy loss,''
  Nucl.\ Phys.\  B {\bf 594}, 371 (2001)
  [arXiv:nucl-th/0006010].
  %%CITATION = NUPHA,B594,371;%%

%\cite{Gyulassy:2002yv}
\bibitem{Gyulassy:2002yv}
  M.~Gyulassy, P.~Levai and I.~Vitev,
  %``Reaction operator approach to multiple elastic scatterings,''
  Phys.\ Rev.\  D {\bf 66}, 014005 (2002)
  [arXiv:nucl-th/0201078].
  %%CITATION = PHRVA,D66,014005;%%


%\cite{Gyulassy:2001kr}
\bibitem{Gyulassy:2001kr}
  M.~Gyulassy, I.~Vitev, X.~N.~Wang and P.~Huovinen,
  %``Transverse Expansion and High $p_T$ Azimuthal Asymmetry at RHIC,''
  Phys.\ Lett.\  B {\bf 526}, 301 (2002)
  [arXiv:nucl-th/0109063].
  %%CITATION = PHLTA,B526,301;%%

% Estimates 100 \rho_A
%\cite{Gyulassy:2001nm}
\bibitem{Gyulassy:2001nm}
  M.~Gyulassy, P.~Levai and I.~Vitev,
  %``Jet tomography of Au + Au reactions including multi-gluon fluctuations,''
  Phys.\ Lett.\  B {\bf 538}, 282 (2002)
  [arXiv:nucl-th/0112071].
  %%CITATION = PHLTA,B538,282;%%



%\cite{Salgado:2002cd}
\bibitem{Salgado:2002cd}
  C.~A.~Salgado and U.~A.~Wiedemann,
  %``A dynamical scaling law for jet tomography,''
  Phys.\ Rev.\ Lett.\  {\bf 89}, 092303 (2002)
  [arXiv:hep-ph/0204221].
  %%CITATION = PRLTA,89,092303;%%

%\cite{Wang:2002ri}
\bibitem{Wang:2002ri}
  E.~Wang and X.~N.~Wang,
  %``Jet tomography of dense and nuclear matter,''
  Phys.\ Rev.\ Lett.\  {\bf 89}, 162301 (2002)
  [arXiv:hep-ph/0202105].
  %%CITATION = PRLTA,89,162301;%%


%\cite{Gyulassy:2003mc}
%\bibitem{Gyulassy:2003mc}
%  M.~Gyulassy, I.~Vitev, X.~N.~Wang and B.~W.~Zhang,
%  %``Jet quenching and radiative energy loss in dense nuclear matter,''
%  arXiv:nucl-th/0302077.
%  %%CITATION = NUCL-TH/0302077;%%


%\cite{Kharzeev:2001gp}
\bibitem{Kharzeev:2001gp}
  D.~Kharzeev and E.~Levin,
  %``Manifestations of high density QCD in the first RHIC data,''
  Phys.\ Lett.\  B {\bf 523}, 79 (2001)
  [arXiv:nucl-th/0108006].
  %%CITATION = PHLTA,B523,79;%%


%\cite{Adler:2002tq}
\bibitem{Adler:2002tq}
  C.~Adler {\it et al.}  [STAR Collaboration],
  %``Disappearance of back-to-back high p(T) hadron correlations in central Au +
  %Au collisions at s(NN)**(1/2) = 200-GeV,''
  Phys.\ Rev.\ Lett.\  {\bf 90}, 082302 (2003)
  [arXiv:nucl-ex/0210033].
  %%CITATION = PRLTA,90,082302;%%

%\cite{Adler:2003pb}
\bibitem{Adler:2003pb}
  S.~S.~Adler {\it et al.}  [PHENIX Collaboration],
  %``Mid-rapidity neutral pion production in proton proton collisions at
  %s**(1/2) = 200-GeV,''
  Phys.\ Rev.\ Lett.\  {\bf 91}, 241803 (2003)
  [arXiv:hep-ex/0304038].
  %%CITATION = PRLTA,91,241803;%%

%\cite{Back:2003ns}
\bibitem{Back:2003ns}
  B.~B.~Back {\it et al.}  [PHOBOS Collaboration],
  %``Centrality dependence of charged hadron transverse momentum spectra in  d +
  %Au collisions at s(NN)**(1/2) = 200-GeV,''
  Phys.\ Rev.\ Lett.\  {\bf 91}, 072302 (2003)
  [arXiv:nucl-ex/0306025].
  %%CITATION = PRLTA,91,072302;%%


%\cite{Adler:2003ii}
\bibitem{Adler:2003ii}
  S.~S.~Adler {\it et al.}  [PHENIX Collaboration],
  %``Absence of suppression in particle production at large transverse  momentum
  %in s(NN)**(1/2) = 200-GeV d + Au collisions,''
  Phys.\ Rev.\ Lett.\  {\bf 91}, 072303 (2003)
  [arXiv:nucl-ex/0306021].
  %%CITATION = PRLTA,91,072303;%%

%\cite{Adams:2003im}
\bibitem{Adams:2003im}
  J.~Adams {\it et al.}  [STAR Collaboration],
  %``Evidence from d + Au measurements for final-state suppression of high  p(T)
  %hadrons in Au + Au collisions at RHIC,''
  Phys.\ Rev.\ Lett.\  {\bf 91}, 072304 (2003)
  [arXiv:nucl-ex/0306024].
  %%CITATION = PRLTA,91,072304;%%

%\cite{Arsene:2003yk}
\bibitem{Arsene:2003yk}
  I.~Arsene {\it et al.}  [BRAHMS Collaboration],
  %``Transverse momentum spectra in Au + Au and d + Au collisions at
  %s(NN)**(1/2) = 200-GeV and the pseudorapidity dependence of high p(T)
  %suppression,''
  Phys.\ Rev.\ Lett.\  {\bf 91}, 072305 (2003)
  [arXiv:nucl-ex/0307003].
  %%CITATION = PRLTA,91,072305;%%

%\cite{Bass:1999zq}
\bibitem{Bass:1999zq}
  S.~A.~Bass {\it et al.},
  %``Last call for RHIC predictions,''
  Nucl.\ Phys.\  A {\bf 661}, 205 (1999)
  [arXiv:nucl-th/9907090].
  %%CITATION = NUPHA,A661,205;%%

%\cite{Back:2000gw}
\bibitem{Back:2000gw}
  B.~B.~Back {\it et al.}  [PHOBOS Collaboration],
  %``Charged particle multiplicity near mid-rapidity in central Au + Au
  %collisions at s**(1/2) = 56-A/GeV and 130-A/GeV,''
  Phys.\ Rev.\ Lett.\  {\bf 85}, 3100 (2000)
  [arXiv:hep-ex/0007036].
  %%CITATION = PRLTA,85,3100;%%

%\cite{Adcox:2000sp}
\bibitem{Adcox:2000sp}
  K.~Adcox {\it et al.}  [PHENIX Collaboration],
  %``Centrality dependence of charged particle multiplicity in Au Au  collisions
  %at s(N N)**(1/2) = 130-GeV,''
  Phys.\ Rev.\ Lett.\  {\bf 86}, 3500 (2001)
  [arXiv:nucl-ex/0012008].
  %%CITATION = PRLTA,86,3500;%%

%\cite{Adler:2004zn}
\bibitem{Adler:2004zn}
  S.~S.~Adler {\it et al.}  [PHENIX Collaboration],
  %``Systematic studies of the centrality and s(NN)**(1/2) dependence of
  %dE(T)/d mu and d N(ch)/d mu in heavy ion collisions at mid-rapidity,''
  Phys.\ Rev.\  C {\bf 71}, 034908 (2005)
  [Erratum-ibid.\  C {\bf 71}, 049901 (2005)]
  [arXiv:nucl-ex/0409015].
  %%CITATION = PHRVA,C71,034908;%%

%\cite{Kharzeev:2001yq}
\bibitem{Kharzeev:2001yq}
  D.~Kharzeev, E.~Levin and M.~Nardi,
  %``The onset of classical QCD dynamics in relativistic heavy ion
  %collisions,''
  Phys.\ Rev.\  C {\bf 71}, 054903 (2005)
  [arXiv:hep-ph/0111315].
  %%CITATION = PHRVA,C71,054903;%%


%\cite{Back:2001ae}
\bibitem{Back:2001ae}
  B.~B.~Back {\it et al.}  [PHOBOS Collaboration],
  %``Energy dependence of particle multiplicities in central Au + Au
  %collisions,''
  Phys.\ Rev.\ Lett.\  {\bf 88}, 022302 (2002)
  [arXiv:nucl-ex/0108009].
  %%CITATION = PRLTA,88,022302;%%

%\cite{Hirano:2004en}
\bibitem{Hirano:2004en}
  T.~Hirano and Y.~Nara,
  %``Hydrodynamic afterburner for the color glass condensate and the parton
  %energy loss,''
  Nucl.\ Phys.\  A {\bf 743}, 305 (2004)
  [arXiv:nucl-th/0404039].
  %%CITATION = NUPHA,A743,305;%%

%\cite{Adcox:2001mf}
\bibitem{Adcox:2001mf}
  K.~Adcox {\it et al.}  [PHENIX Collaboration],
  %``Centrality dependence of pi+-, K+-, p and anti-p production from
  %s(NN)**(1/2) = 130-GeV Au + Au collisions at RHIC,''
  Phys.\ Rev.\ Lett.\  {\bf 88}, 242301 (2002)
  [arXiv:nucl-ex/0112006].
  %%CITATION = PRLTA,88,242301;%%

%\cite{Hwa:2002tu}
\bibitem{Hwa:2002tu}
  R.~C.~Hwa and C.~B.~Yang,
  %``Scaling behavior at high p(T) and the p/pi ratio,''
  Phys.\ Rev.\  C {\bf 67}, 034902 (2003)
  [arXiv:nucl-th/0211010].
  %%CITATION = PHRVA,C67,034902;%%


%\cite{Fries:2003vb}
\bibitem{Fries:2003vb}
  R.~J.~Fries, B.~Muller, C.~Nonaka and S.~A.~Bass,
  %``Hadronization in heavy ion collisions: Recombination and fragmentation  of
  %partons,''
  Phys.\ Rev.\ Lett.\  {\bf 90}, 202303 (2003)
  [arXiv:nucl-th/0301087].
  %%CITATION = PRLTA,90,202303;%%

%\cite{Greco:2003xt}
\bibitem{Greco:2003xt}
  V.~Greco, C.~M.~Ko and P.~Levai,
  %``Parton coalescence and antiproton/pion anomaly at RHIC,''
  Phys.\ Rev.\ Lett.\  {\bf 90}, 202302 (2003)
  [arXiv:nucl-th/0301093].
  %%CITATION = PRLTA,90,202302;%%

%\cite{Greco:2003mm}
\bibitem{Greco:2003mm}
  V.~Greco, C.~M.~Ko and P.~Levai,
  %``Parton coalescence at RHIC,''
  Phys.\ Rev.\  C {\bf 68}, 034904 (2003)
  [arXiv:nucl-th/0305024].
  %%CITATION = PHRVA,C68,034904;%%


%\cite{Fries:2003kq}
\bibitem{Fries:2003kq}
  R.~J.~Fries, B.~Muller, C.~Nonaka and S.~A.~Bass,
  %``Hadron production in heavy ion collisions: Fragmentation and  recombination
  %from a dense parton phase,''
  Phys.\ Rev.\  C {\bf 68}, 044902 (2003)
  [arXiv:nucl-th/0306027].
  %%CITATION = PHRVA,C68,044902;%%

%\cite{Muller:2006ee}
\bibitem{Muller:2006ee}
  B.~Muller and J.~L.~Nagle,
  %``Results from the Relativistic Heavy Ion Collider,''
  Ann.\ Rev.\ Nucl.\ Part.\ Sci.\  {\bf 56}, 93 (2006)
  [arXiv:nucl-th/0602029].
  %%CITATION = ARNUA,56,93;%%

%\cite{Adler:2005ig}
\bibitem{Adler:2005ig}
  S.~S.~Adler {\it et al.}  [PHENIX Collaboration],
  %``Centrality dependence of direct photon production in s(NN)**(1/2) =
  %200-GeV Au + Au collisions,''
  Phys.\ Rev.\ Lett.\  {\bf 94}, 232301 (2005)
  [arXiv:nucl-ex/0503003].
  %%CITATION = PRLTA,94,232301;%%


%\cite{Turbide:2005fk}
\bibitem{Turbide:2005fk}
  S.~Turbide, C.~Gale, S.~Jeon and G.~D.~Moore,
  %``Energy loss of leading hadrons and direct photon production in evolving
  %quark-gluon plasma,''
  Phys.\ Rev.\  C {\bf 72}, 014906 (2005)
  [arXiv:hep-ph/0502248].
  %%CITATION = PHRVA,C72,014906;%%


%\cite{Arleo:2006xb}
\bibitem{Arleo:2006xb}
  F.~Arleo,
  %``Hard pion and prompt photon at RHIC, from single to double inclusive
  %production,''
  JHEP {\bf 0609}, 015 (2006)
  [arXiv:hep-ph/0601075].
  %%CITATION = JHEPA,0609,015;%%

%\cite{Jin:2007by}
\bibitem{Jin:2007by}
  J.~Jin  [PHENIX Collaboration],
  %``PHENIX measurement of high-p(T) hadron hadron and photon hadron azimuthal
  %correlations,''
  J.\ Phys.\ G {\bf 34}, S813 (2007)
  [arXiv:0705.0842 [nucl-ex]].
  %%CITATION = JPHGB,G34,S813;%%

%\cite{Chattopadhyay:2007in}
\bibitem{Chattopadhyay:2007in}
  S.~Chattopadhyay  [STAR Collaboration],
  %``Correlation between photons and charged particles in d Au collisions at
  %RHIC,''
  Nucl.\ Phys.\  A {\bf 783} (2007) 591.
  %%CITATION = NUPHA,A783,591;%%

%\cite{Wang:1996yh}
\bibitem{Wang:1996yh}
  X.~N.~Wang, Z.~Huang and I.~Sarcevic,
  %``Jet quenching in the opposite direction of a tagged photon in  high-energy
  %heavy-ion collisions,''
  Phys.\ Rev.\ Lett.\  {\bf 77}, 231 (1996)
  [arXiv:hep-ph/9605213].
  %%CITATION = PRLTA,77,231;%%


\bibitem{Weinberg}
S.~Weinberg, {\em Gravitation and Cosmology}, John Wiley and Sons,
1972.

%\cite{Csanad:2006sp}
\bibitem{Csanad:2006sp}
  See references in
  M.~Csanad, T.~Csorgo, R.~A.~Lacey and B.~Lorstad,
  %``Universal scaling of the elliptic flow at RHIC,''
  arXiv:nucl-th/0605044.
  %%CITATION = NUCL-TH/0605044;%%

%\cite{Adare:2006ti}
\bibitem{Adare:2006ti}
  A.~Adare {\it et al.}  [PHENIX Collaboration],
  %``Scaling properties of azimuthal anisotropy in Au + Au and Cu + Cu
  %collisions at s(NN)**(1/2) = 200-GeV,''
  Phys.\ Rev.\ Lett.\  {\bf 98}, 162301 (2007)
  [arXiv:nucl-ex/0608033].
  %%CITATION = PRLTA,98,162301;%%

%\cite{Adams:2003am}
\bibitem{Adams:2003am}
  J.~Adams {\it et al.}  [STAR Collaboration],
  %``Particle dependence of azimuthal anisotropy and nuclear modification of
  %particle production at moderate p(T) in Au + Au collisions at  s(NN)**(1/2) =
  %200-GeV,''
  Phys.\ Rev.\ Lett.\  {\bf 92}, 052302 (2004)
  [arXiv:nucl-ex/0306007].
  %%CITATION = PRLTA,92,052302;%%

%\cite{Adams:2005zg}
\bibitem{Adams:2005zg}
  J.~Adams {\it et al.}  [STAR Collaboration],
  %``Multi-strange baryon elliptic flow in Au + Au collisions at  s(NN)**(1/2) =
  %200-GeV,''
  Phys.\ Rev.\ Lett.\  {\bf 95}, 122301 (2005)
  [arXiv:nucl-ex/0504022].
  %%CITATION = PRLTA,95,122301;%%

%\cite{Belenkij:1956cd}
\bibitem{Belenkij:1956cd}
  S.~Z.~Belenkij and L.~D.~Landau,
  %``Hydrodynamic theory of multiple production of particles,''
  Nuovo Cim.\ Suppl.\  {\bf 3S10}, 15 (1956)
  [Usp.\ Fiz.\ Nauk {\bf 56}, 309 (1955)].
  %%CITATION = UFNAA,56,309;%%

%\cite{Kovtun:2004de}
\bibitem{Kovtun:2004de}
  P.~Kovtun, D.~T.~Son and A.~O.~Starinets,
  %``Viscosity in strongly interacting quantum field theories from black hole
  %physics,''
  Phys.\ Rev.\ Lett.\  {\bf 94}, 111601 (2005)
  [arXiv:hep-th/0405231].
  %%CITATION = PRLTA,94,111601;%%


%\cite{Danielewicz:1984ww}
\bibitem{Danielewicz:1984ww}
  P.~Danielewicz and M.~Gyulassy,
  %``Dissipative Phenomena In Quark Gluon Plasmas,''
  Phys.\ Rev.\  D {\bf 31}, 53 (1985).
  %%CITATION = PHRVA,D31,53;%%

%\cite{Hirano:2005wx}
\bibitem{Hirano:2005wx}
  T.~Hirano and M.~Gyulassy,
  %``Perfect fluidity of the quark gluon plasma core as seen through its
  %dissipative hadronic corona,''
  Nucl.\ Phys.\  A {\bf 769}, 71 (2006)
  [arXiv:nucl-th/0506049].
  %%CITATION = NUPHA,A769,71;%%


%\cite{Maldacena:1997re}
\bibitem{Maldacena:1997re}
  J.~M.~Maldacena,
  %``The large N limit of superconformal field theories and supergravity,''
  Adv.\ Theor.\ Math.\ Phys.\  {\bf 2}, 231 (1998)
  [Int.\ J.\ Theor.\ Phys.\  {\bf 38}, 1113 (1999)]
  [arXiv:hep-th/9711200].
  %%CITATION = IJTPB,38,1113;%%


%\cite{Policastro:2001yc}
\bibitem{Policastro:2001yc}
  G.~Policastro, D.~T.~Son and A.~O.~Starinets,
  %``The shear viscosity of strongly coupled N = 4 supersymmetric Yang-Mills
  %plasma,''
  Phys.\ Rev.\ Lett.\  {\bf 87}, 081601 (2001)
  [arXiv:hep-th/0104066].
  %%CITATION = PRLTA,87,081601;%%

%\cite{Gubser:2006bz}
\bibitem{Gubser:2006bz}
  S.~S.~Gubser,
  %``Drag force in AdS/CFT,''
  Phys.\ Rev.\  D {\bf 74}, 126005 (2006)
  [arXiv:hep-th/0605182].
  %%CITATION = PHRVA,D74,126005;%%

%\cite{Liu:2006he}
\bibitem{Liu:2006he}
  H.~Liu, K.~Rajagopal and U.~A.~Wiedemann,
  %``Wilson loops in heavy ion collisions and their calculation in AdS/CFT,''
  JHEP {\bf 0703}, 066 (2007)
  [arXiv:hep-ph/0612168].
  %%CITATION = JHEPA,0703,066;%%

%\cite{Teaney:2003kp}
\bibitem{Teaney:2003kp}
  D.~Teaney,
  %``Effect of shear viscosity on spectra, elliptic flow, and Hanbury
  %Brown-Twiss radii,''
  Phys.\ Rev.\  C {\bf 68}, 034913 (2003)
  [arXiv:nucl-th/0301099].
  %%CITATION = PHRVA,C68,034913;%%

%\cite{Shuryak:2003xe}
\bibitem{Shuryak:2003xe}
  E.~Shuryak,
  %``Why does the quark gluon plasma at RHIC behave as a nearly ideal fluid?,''
  Prog.\ Part.\ Nucl.\ Phys.\  {\bf 53}, 273 (2004)
  [arXiv:hep-ph/0312227].
  %%CITATION = PPNPD,53,273;%%

%\cite{Lacey:2006bc}
\bibitem{Lacey:2006bc}
  R.~A.~Lacey {\it et al.},
  %``Has the QCD critical point been signaled by observations at RHIC?,''
  Phys.\ Rev.\ Lett.\  {\bf 98}, 092301 (2007)
  [arXiv:nucl-ex/0609025].
  %%CITATION = PRLTA,98,092301;%%

%\cite{Drescher:2007cd}
\bibitem{Drescher:2007cd}
  H.~J.~Drescher, A.~Dumitru, C.~Gombeaud and J.~Y.~Ollitrault,
  %``The centrality dependence of elliptic flow, the hydrodynamic limit, and
  %the viscosity of hot QCD,''
  Phys.\ Rev.\  C {\bf 76}, 024905 (2007)
  [arXiv:0704.3553 [nucl-th]].
  %%CITATION = PHRVA,C76,024905;%%

%\cite{Gavin:2006xd}
\bibitem{Gavin:2006xd}
  S.~Gavin and M.~Abdel-Aziz,
  %``Measuring Shear Viscosity Using Transverse Momentum Correlations in
  %Relativistic Nuclear Collisions,''
  Phys.\ Rev.\ Lett.\  {\bf 97}, 162302 (2006)
  [arXiv:nucl-th/0606061].
  %%CITATION = PRLTA,97,162302;%%

%\cite{Adare:2006nq}
\bibitem{Adare:2006nq}
  A.~Adare {\it et al.}  [PHENIX Collaboration],
  %``Energy loss and flow of heavy quarks in Au + Au collisions at  s(NN)**(1/2)
  %= 200-GeV,''
  Phys.\ Rev.\ Lett.\  {\bf 98}, 172301 (2007)
  [arXiv:nucl-ex/0611018].
  %%CITATION = PRLTA,98,172301;%%

%\cite{Schafer:2007waz}
\bibitem{Schafer:2007waz}
  T.~Sch\"{a}fer,
  %``The Shear Viscosity to Entropy Density Ratio of Trapped Fermions in the Unitarity Limit,''
  arXiv:cond-mat/0701251v3 .

%\cite{Shuryak:2006ii}
\bibitem{Shuryak:2006ii}
  E.~Shuryak,
  %``The conical flow from quenched jets in sQGP,''
  Nucl.\ Phys.\  A {\bf 783}, 31 (2007)
  [arXiv:nucl-th/0609013].
  %%CITATION = NUPHA,A783,31;%%



%\cite{Stoecker:2004qu}
\bibitem{Stoecker:2004qu}
  H.~Stoecker,
  %``Collective Flow signals the Quark Gluon Plasma,''
  Nucl.\ Phys.\  A {\bf 750}, 121 (2005)
  [arXiv:nucl-th/0406018].
  %%CITATION = NUPHA,A750,121;%%


%\cite{CasalderreySolana:2004qm}
\bibitem{CasalderreySolana:2004qm}
  J.~Casalderrey-Solana, E.~V.~Shuryak and D.~Teaney,
  %``Conical flow induced by quenched QCD jets,''
  J.\ Phys.\ Conf.\ Ser.\  {\bf 27}, 22 (2005)
  [Nucl.\ Phys.\  A {\bf 774}, 577 (2006)]
  [arXiv:hep-ph/0411315].
  %%CITATION = NUPHA,A774,577;%%

%\cite{Adler:2005ee}
\bibitem{Adler:2005ee}
  S.~S.~Adler {\it et al.}  [PHENIX Collaboration],
  %``Modifications to di-jet hadron pair correlations in Au + Au collisions  at
  %s(NN)**(1/2) = 200-GeV,''
  Phys.\ Rev.\ Lett.\  {\bf 97}, 052301 (2006)
  [arXiv:nucl-ex/0507004].
  %%CITATION = PRLTA,97,052301;%%

%\cite{Adare:2006nr}
\bibitem{Adare:2006nr}
  A.~Adare {\it et al.}  [PHENIX Collaboration],
  %``System size and energy dependence of jet-induced hadron pair  correlation
  %shapes in Cu + Cu and Au + Au collisions at s(NN)**(1/2) =  200-GeV and
  %62.4-GeV,''
  Phys.\ Rev.\ Lett.\  {\bf 98}, 232302 (2007)
  [arXiv:nucl-ex/0611019].
  %%CITATION = PRLTA,98,232302;%%

%
% In the data, not the statements
%
%\cite{Horner:2007gt}
\bibitem{Horner:2007gt}
  M.~J.~Horner  [STAR Collaboration],
  %``Low- and intermediate-p(T) di-hadron distributions in Au + Au collisions at
  %s(NN)**(1/2) = 200-GeV from STAR,''
  J.\ Phys.\ G {\bf 34}, S995 (2007)
  [arXiv:nucl-ex/0701069].
  %%CITATION = JPHGB,G34,S995;%%

%\cite{Ulery:2007zb}
\bibitem{Ulery:2007zb}
  J.~G.~Ulery  [STAR Collaboration],
  %``Are There Mach Cones in Heavy Ion Collisions? Three-Particle   Correlations
  %from STAR,''
  arXiv:0704.0224 [nucl-ex].
  %%CITATION = ARXIV:0704.0224;%%

%\cite{Jia:2007sf}
\bibitem{Jia:2007sf}
  J.~Jia  [for the PHENIX Collaboration],
  %``Mapping out the Jet correlation landscape: Jet quenching and Medium
  %response,''
  arXiv:0705.3060 [nucl-ex].
  %%CITATION = ARXIV:0705.3060;%%

%\cite{Gubser:2007ni}
\bibitem{Gubser:2007ni}
  See S.~S.~Gubser and A.~Yarom,
  %``Universality of the diffusion wake in the gauge-string duality,''
  arXiv:0709.1089 [hep-th], and references therein.
  %%CITATION = ARXIV:0709.1089;%%

%\cite{Adams:2004bi}
\bibitem{Adams:2004bi}
  J.~Adams {\it et al.}  [STAR Collaboration],
  %``Azimuthal anisotropy in Au + Au collisions at s(NN)**(1/2) = 200-GeV,''
  Phys.\ Rev.\  C {\bf 72}, 014904 (2005)
  [arXiv:nucl-ex/0409033].
  %%CITATION = PHRVA,C72,014904;%%

%\cite{Afanasiev:2007tv}
\bibitem{Afanasiev:2007tv}
  S.~Afanasiev {\it et al.}  [PHENIX Collaboration],
  %``Elliptic flow for $\phi$ mesons and (anti)deuterons in Au+Au collisions at
  %$\sqrt{s_{NN}}$ = 200 GeV,''
  arXiv:nucl-ex/0703024.
  %%CITATION = NUCL-EX/0703024;%%

%\cite{Abelev:2007rw}
\bibitem{Abelev:2007rw}
  B.~I.~Abelev {\it et al.}  [STAR Collaboration],
  %``Partonic flow and Phi-meson production in Au + Au collisions at
  %s(NN)**(1/2) = 200-GeV,''
  arXiv:nucl-ex/0703033.
  %%CITATION = NUCL-EX/0703033;%%


\bibitem{QP}
  The author would like to thank
  M.~Csan\'{a}d, T.~Cs\"{o}rg\H{o}, B.~M\"{u}ller, J.~Nagle, K.~Rajagopal, T.~Sch\"{a}fer, and D.~Son
  for useful discussions on many aspects of quasi-particles.

%\cite{Song:2007fn}
\bibitem{Song:2007fn}
  H.~Song and U.~W.~Heinz,
  %``Suppression of elliptic flow in a minimally viscous quark-gluon plasma,''
  arXiv:0709.0742 [nucl-th].
  %%CITATION = ARXIV:0709.0742;%%


%\cite{Antinori:2007dna}
\bibitem{Antinori:2007dna}
  F.~Antinori  [ALICE Collaboration],
  %``Heavy-ion physics with ALICE,''
  J.\ Phys.\ G {\bf 34}, S511 (2007)
  [arXiv:nucl-ex/0702013].
  %%CITATION = JPHGB,G34,S511;%%

%\cite{Betts:2007zz}
\bibitem{Betts:2007zz}
  R.~R.~Betts,
  %``Heavy-Ion Physics with CMS,''
  J.\ Phys.\ G {\bf 34}, S519 (2007).
  %%CITATION = JPHGB,G34,S519;%%

%\cite{Steinberg:2007nm}
\bibitem{Steinberg:2007nm}
  P.~Steinberg  [ATLAS Collaboration],
  %``Heavy ion physics at the LHC with the ATLAS detector,''
  J.\ Phys.\ G {\bf 34}, S527 (2007)
  [arXiv:0705.0382 [nucl-ex]].
  %%CITATION = JPHGB,G34,S527;%%

%\cite{Stephanov:1998dy}
\bibitem{Stephanov:1998dy}
  M.~A.~Stephanov, K.~Rajagopal and E.~V.~Shuryak,
  %``Signatures of the tricritical point in {QCD},''
  Phys.\ Rev.\ Lett.\  {\bf 81}, 4816 (1998)
  [arXiv:hep-ph/9806219].
  %%CITATION = PRLTA,81,4816;%%


%\cite{Stephans:2006tg}
\bibitem{Stephans:2006tg}
  G.~S.~F.~Stephans,
  %``critRHIC: The RHIC low energy program,''
  J.\ Phys.\ G {\bf 32}, S447 (2006)
  [arXiv:nucl-ex/0607030].
  %%CITATION = JPHGB,G32,S447;%%

%\cite{Csernai:2006zz}
\bibitem{Csernai:2006zz}
  L.~P.~Csernai, J.~I.~Kapusta and L.~D.~McLerran,
  %``On the strongly-interacting low-viscosity matter created in relativistic
  %nuclear collisions,''
  Phys.\ Rev.\ Lett.\  {\bf 97}, 152303 (2006)
  [arXiv:nucl-th/0604032].
  %%CITATION = PRLTA,97,152303;%%

%\cite{Henning:2004vk}
\bibitem{Henning:2004vk}
  W.~F.~Henning,
  %``FAIR - An International Accelerator Facility for Research with Ions and
  %Antiprotons,''
  AIP Conf.\ Proc.\  {\bf 773}, 3 (2005).
  %%CITATION = APCPC,773,3;%%

% notes:

% \bibitem{label2}
% If there is a note, it should come last:
% \bibitem{label3} \note
% \end{subbibitems}


\end{thebibliography}
\end{document}